\let\reset@font\empty
\def\fileversion{v2.5}
\def\filedate{4 October 1993}

\newdimen\@bls                    % \@b(ase)l(ine)s(kip)
\@bls=\baselineskip               % \@bls ~= \baselineskip for \normalsize
\advance\@bls -1ex                % (fudge term)
\newdimen\@eps                    %
\def\section{\@startsection{section}{1}{\z@}
  {1.5\@bls plus 0.5\@bls}{1\@bls}{\normalsize\bf}}
\def\subsection{\@startsection{subsection}{2}{\z@}
  {1\@bls plus 0.25\@bls}{\@eps}{\normalsize\bf}}
\def\subsubsection{\@startsection{subsubsection}{3}{\z@}
  {1\@bls plus 0.25\@bls}{\@eps}{\normalsize\bf}}
\def\paragraph{\@startsection{paragraph}{4}{\parindent}
  {1\@bls plus 0.25\@bls}{0.5em}{\normalsize\bf}}
\def\subparagraph{\@startsection{subparagraph}{4}{\parindent}
  {1\@bls plus 0.25\@bls}{0.5em}{\normalsize\bf}}

\def\@sect#1#2#3#4#5#6[#7]#8{\ifnum #2>\c@secnumdepth
  \def\@svsec{}\else
  \refstepcounter{#1}\edef\@svsec{\csname the#1\endcsname.\hskip0.5em}\fi
  \@tempskipa #5\relax
  \ifdim \@tempskipa>\z@
    \begingroup
      #6\relax
      \@hangfrom{\hskip #3\relax\@svsec}{\interlinepenalty \@M #8\par}%
    \endgroup
    \csname #1mark\endcsname{#7}\addcontentsline
      {toc}{#1}{\ifnum #2>\c@secnumdepth \else
        \protect\numberline{\csname the#1\endcsname}\fi #7}%
  \else
    \def\@svsechd{#6\hskip #3\@svsec #8\csname #1mark\endcsname
      {#7}\addcontentsline{toc}{#1}{\ifnum #2>\c@secnumdepth \else
        \protect\numberline{\csname the#1\endcsname}\fi #7}}%
  \fi \@xsect{#5}}

\long\def\@makefigurecaption#1#2{\vskip 10mm #1. #2\par}

\long\def\@maketablecaption#1#2{\hbox to \hsize{\parbox[t]{\hsize}
  {#1 \\ #2}}\vskip 0.3ex}

\def\fnum@figure{Figure \thefigure}
\def\figure{\let\@makecaption\@makefigurecaption \@float{figure}}
\@namedef{figure*}{\let\@makecaption\@makefigurecaption \@dblfloat{figure}}

\def\table{\let\@makecaption\@maketablecaption \@float{table}}
\@namedef{table*}{\let\@makecaption\@maketablecaption \@dblfloat{table}}

\textfloatsep=\floatsep           % Space between main text and floats
\intextsep=\floatsep              % Space between in-text figures and
\long\def\@makefntext#1{\parindent 1em\noindent\hbox{${}^{\@thefnmark}$}#1}

\def\maketitle{\begingroup        % Initialize generation of front-matter
    \def\thefootnote{\fnsymbol{footnote}}%
    \newpage \global\@topnum\z@
    \@maketitle \@thanks
  \endgroup
  \let\maketitle\relax \let\@maketitle\relax
  \gdef\@thanks{}\let\thanks\relax
  \gdef\@address{}\gdef\@author{}\gdef\@title{}\let\address\relax}

\def\justify@on{\let\\=\@normalcr
  \leftskip\z@ \@rightskip\z@ \rightskip\@rightskip}

\newbox\fm@box                    % Box to capture front-matter in

\def\@maketitle{%                 % Actual formatting of \maketitle
  \global\setbox\fm@box=\vbox\bgroup
    \vskip 8mm                    % 930715: 8mm white space above title
    \raggedright                  % Front-matter text is ragged right
    \hyphenpenalty\@M             % and is not hyphenated.
    {\Large \@title \par}         % Title set in larger font.
    \vskip\@bls                   % One line of vertical space after title.
    {\normalsize                  % each author set in the normal
     \@author \par}               % typeface size
    \vskip\@bls                   % One line of vertical space after author(s).
    \@address                     % all addresses
  \egroup
  \twocolumn[%                    % Front-matter text is over 2 columns.
    \unvbox\fm@box                % Unwrap contents of front-matter box
    \vskip\@bls                   % add 1 line of vertical space,
    \unvbox\abstract@box          % unwrap contents of abstract boxes,
    \vskip 2pc]}                  % and add 2pc of vertical space

\newcounter{address}
\def\theaddress{\alph{address}}
\def\@makeadmark#1{\hbox{$^{\rm #1}$}}

\def\address#1{\addressmark\begingroup
  \xdef\@tempa{\theaddress}\let\\=\relax
  \def\protect{\noexpand\protect\noexpand}\xdef\@address{\@address
  \protect\addresstext{\@tempa}{#1}}\endgroup}
\def\@address{}

\def\addressmark{\stepcounter{address}%
  \xdef\@tempa{\theaddress}\@makeadmark{\@tempa}}

\def\addresstext#1#2{\leavevmode \begingroup
  \raggedright \hyphenpenalty\@M \@makeadmark{#1}#2\par \endgroup
  \vskip\@bls}

\newbox\abstract@box              % Box to capture abstract in

\def\abstract{%
  \global\setbox\abstract@box=\vbox\bgroup
  \small\rm
  \ignorespaces}
\def\endabstract{\par \egroup}

\def\thebibliography#1{\section*{REFERENCES}\list{\arabic{enumi}.}
  {\settowidth\labelwidth{#1.}\leftmargin=1.67em
   \labelsep\leftmargin \advance\labelsep-\labelwidth
   \itemsep\z@ \parsep\z@
   \usecounter{enumi}}\def\makelabel##1{\rlap{##1}\hss}%
   \def\newblock{\hskip 0.11em plus 0.33em minus -0.07em}
   \sloppy \clubpenalty=4000 \widowpenalty=4000 \sfcode`\.=1000\relax}

\newcount\@tempcntc
\def\@citex[#1]#2{\if@filesw\immediate\write\@auxout{\string\citation{#2}}\fi
  \@tempcnta\z@\@tempcntb\m@ne\def\@citea{}\@cite{\@for\@citeb:=#2\do
    {\@ifundefined
       {b@\@citeb}{\@citeo\@tempcntb\m@ne\@citea
        \def\@citea{,\penalty\@m\ }{\bf ?}\@warning
       {Citation `\@citeb' on page \thepage \space undefined}}%
    {\setbox\z@\hbox{\global\@tempcntc0\csname b@\@citeb\endcsname\relax}%
     \ifnum\@tempcntc=\z@ \@citeo\@tempcntb\m@ne
       \@citea\def\@citea{,\penalty\@m}
       \hbox{\csname b@\@citeb\endcsname}%
     \else
      \advance\@tempcntb\@ne
      \ifnum\@tempcntb=\@tempcntc
      \else\advance\@tempcntb\m@ne\@citeo
      \@tempcnta\@tempcntc\@tempcntb\@tempcntc\fi\fi}}\@citeo}{#1}}

\def\@citeo{\ifnum\@tempcnta>\@tempcntb\else\@citea
  \def\@citea{,\penalty\@m}%
  \ifnum\@tempcnta=\@tempcntb\the\@tempcnta\else
   {\advance\@tempcnta\@ne\ifnum\@tempcnta=\@tempcntb \else \def\@citea{--}\fi
    \advance\@tempcnta\m@ne\the\@tempcnta\@citea\the\@tempcntb}\fi\fi}

\def\ps@crcplain{\let\@mkboth\@gobbletwo
     \def\@oddhead{\reset@font{\sl\rightmark}\hfil \rm\thepage}%
     \def\@evenhead{\reset@font\rm \thepage\hfil\sl\leftmark}%
     \let\@oddfoot\@empty
     \let\@evenfoot\@oddfoot}

\ps@crcplain                    % modified 'plain' page style

\documentstyle[twoside,fleqn,espcrc2]{article}

\newcommand{\AmS}{{\protect\the\textfont2
  A\kern-.1667em\lower.5ex\hbox{M}\kern-.125emS}}

\title{Analytical calculation of critical temperatures in some
spin systems \thanks{Presented by J. Wosiek}
\thanks{Supported by the KBN grant No.
 PB 0428/P3.} }

\author{Z. Burda\address{Department of Physics,
        University of Bielefeld, \\
        P.O. Box 8640, D-4800 Bielefeld 1, Germany}%
        \thanks{Fellow of the Alexander von Humboldt Foundation.}
        \thanks{On leave from Jagellonian University, Cracow, Poland.}
        and
        J. Wosiek\address{Institute of Computer Science,
        Jagellonian University, \\
        Reymonta 4, Cracow, Poland}}

\begin{document}

\begin{abstract}
A new method for locating analytically critical temperatures is discussed.
It is exact for selfdual systems. When applied the two coupled layers
of Ising spins it deviates from our preliminary Monte Carlo estimates
by 1.5 standard deviations. It predicts critical temperature of the three
dimensional Ising model in terms of the solution of the two layer Ising
system.

\end{abstract}

\maketitle

\section{The method - selfdual examples}
One of us has proposed recently to
 determine a critical temperature of discrete spin systems
from the position of the maximum of the normalized moments of the transfer
matrix ${\cal T}$ \cite{ja}. Define the following
``characteristic function'' of the $d$ dimensional system
\begin{equation}
\rho_n(\beta) = \lim_{L\rightarrow\infty} \left( {(Tr {\cal T})^n \over
Tr {\cal T}^n }
\right )^{{1\over L^{d-1}}} ,  \label{rho}
\end{equation}
In the following we will work mainly with $n=2$ and often skip the
subscript $n$.
The rationale behind Eq.(\ref{rho}) is that, even though $\rho$ is not
dominated
by the largest eigenvalue, it is conceivable that {\em all} eigenvalues
$\lambda_{\alpha}$ of
the transfer matrix are sensitive to $\beta_c$. In fact famous solution
of the
Ising model \cite{ons,kauf} shows this property explicitly.
All eigenvalues have extremum at the critical temperatue \cite{huang}. If so,
than any simple function of $\lambda$'s should also have this feature. It is
easy to prove that
a) $\rho(0)=1$, b) $\rho(\infty)=1$, and c) $ \rho(\beta) \ge 1$  \cite{ja}.
Together with earlier arguments this forms the basis of our hypothesis,
\begin{equation}
\beta_{max}=\beta_c. \label{hypo}
\end{equation}
For selfdual systems, Eq.(\ref{hypo}) can be proven if $\rho(\beta)$ has
a single
maximum. Hence our ultimate goal is to test this proposition for
{\em not} selfdual
systems. Let us however first illustrate the simplicity of this  approach
on the example
of two and three state Potts models in two dimensions. Low moments of
$\cal T$
are straightforward to calculate since $Tr {\cal T}^n=Z_n=Tr T_n^L$ is the
partition function of $n$ coupled chains, and consequently $T_n$ is the
``small'' transfer matrix propagating a system of $n$ spins
{\em horizontally}
(see Fig.1).

\begin{figure}[htb]
%\vspace{9pt}
%\framebox[55mm]{\rule[-21mm]{0mm}{43mm}}
\begin{picture}(100,50)(-50,20)
\setlength{\unitlength}{.5pt}
\multiput(20,50)(36,0){4}{\circle*{5}}
\multiput(20,86)(36,0){4}{\circle*{5}}
\multiput(145,68)(6,0){3}{\circle*{2}}
\multiput(170,50)(0,36){2}{\circle*{5}}
\put (20,50) {\line(124,0){124} }
\put (20,86) {\line(124,0){124} }
\put (50,120) { $ T_2$  }
\put(80,125){\vector(1,0){40}}
\multiput(20,50)(36,0){4}{\line(0,36){36} }
\put(170,50) {\line(0,36){36} }
\put(210,45) {${\cal T}$}
\put(214,60){\vector(0,1){20}}
%\put(128,-100){{\Large Fig.1}}
\end{picture}
\caption{Construction of the transfer matrices ${\cal T}$ and $T_2$
for the two dimensional Ising model. Periodic boundary conditions
in both directions are understood.}
\label{fig:goodenenough}
\end{figure}
Therefore the second normalized moment $\rho_2(\beta)=(t_{1max})^2/t_{2max}$,
where $t_{1,2 max}$ are the largest eigenvalues of $T_{1,2}$. The problem
of one and two Potts spins can be easily solved. We obtain \cite{ja}
\begin{equation}
\rho(\beta)={Q_1(x)\over Q_2(x) + \sqrt{Q_3(x)} }, x=\exp{\beta},
\end{equation}
where for the Ising case $(q=2)$
\begin{eqnarray}
Q_1(x)&=&2 x^2 (x+1)^2, \nonumber \\
Q_2(x)&=&(x^2+1)^2, \nonumber \\
Q_3(x)&=&(x^2-1)^4+4 x^2 (x^2+1)^2, \nonumber
\end{eqnarray}
and for $q=3$
\begin{eqnarray}
Q_1(x)&=&2 x^2 (x+2)^2, \nonumber \\
Q_2(x)&=&x^4+3x^2+2x+3, \\
Q_3(x)&=&x^8+2x^6-4x^5+27x^4+28x^3 \nonumber \\
      &+&6x^2+12x+9. \nonumber
\end{eqnarray}
In both cases $\rho(\beta)$ has a single maximum at $x_c=1+\sqrt{q}$,
as expected. We have also checked that Eq.(\ref{hypo}) works for
$q=4,5$ and for the Ising model on a triangular lattice.
\subsection{Two layer Ising model}
Consider a system  of two coupled planes of
Ising spins. Impose periodic boundary conditions in all three directions.
this effectively
 doubles the strengh of interaction between the planes.
The system in not selfdual and to our knowledge
 its critical temperature was never derived.  On the other hand
calculation of characteristic function $\rho(\beta)$
is relatively simple and provides analytic predictions for $\beta_{c}$.
 Transfer matrices $T_{1,2}$
propagate now states of two and four spins respectively, see Fig.2.

\begin{figure}[htb]
%\vspace{9pt}
%\framebox[55mm]{\rule[-21mm]{0mm}{43mm}}
\begin{picture}(100,55)(-50,25)
\setlength{\unitlength}{.6pt}
\multiput(20,50)(36,0){4}{\circle*{5}}
\multiput(20,86)(36,0){4}{\circle*{5}}
\multiput(32,68)(36,0){4}{\circle*{5}}
\multiput(32,104)(36,0){4}{\circle*{5}}
\multiput(150,76)(6,0) {3}{\circle*{2}}
\multiput(170,50)(0,36){2}{\circle*{5}}
\multiput(182,68)(0,36){2}{\circle*{5}}
\put (20,50) {\line(124,0){124} }
\put (20,86) {\line(124,0){124} }
\multiput(32,68)(0,36){2}{\line(1,0){124}}
\put (50,138) { $ T_2$  }
\put(80,143){\vector(1,0){40}}
\multiput(20,50)(36,0){4}{\line(0,1){36}}
\multiput(20,50)(36,0){4}{\line(2,3){12}}
\multiput(20,86)(36,0){4}{\line(2,3){12}}
\multiput(32,68)(36,0){4}{\line(0,1){36}}
\multiput(170,50)(12,18){2}{\line(0,1){36}}
\multiput(170,50)(0,36){2}{\line(2,3){12}}
\put(210,45) {${\cal T}$}
\put(214,60){\vector(0,1){20}}
\end{picture}
\caption{The same as Fig.1 but for the two layer Ising model.}
\label{fig:largenenough}
\end{figure}
The largest eigenvalue of $T_1$ is
\begin{equation}
t_1^{max}={x^2\over 2}(x^4+2+x^{-4}+\sqrt{x^8+14+x^{-8}}). \label{t1}
\end{equation}
$T_2$ conserves
the $U$-parity, $[T_2,U]=0, U=\prod_{i=1}^4 \sigma_i^x$.
This reduces the problem to diagonalization of two $8\times 8$ matrices.
The largest eigenvalue
belongs to the $U=+1$ sector. Final expression reads
\begin{eqnarray}
t_2^{max}&=&{(1+x^4)^2 \over 4 x^{12}} Q_1
  +  {1+x^8 \over 4x^{12}}\sqrt{Q_2} \nonumber \\
 &+& {1+x^4 \over 2\sqrt{2} x^{12}} \sqrt{Q_3+(1+x^8) Q_1 \sqrt{Q_2} },
  \\
\label{dlay}
Q_1(x)&=&x^{16}-2x^{12}+6x^8-2x^4+1, \nonumber \\
Q_2(x)&=&x^{32}-4x^{24}+70x^{16}-4x^8+1,\nonumber  \\
Q_3(x)&=&x^{40}-2x^{36}+5x^{32}+26x^{24}+4x^{20} \nonumber \\
 & & + 26x^{16}+5x^8-2x^4+1. \nonumber
\end{eqnarray}
Resulting characteristic function $\rho(\beta)$
has the single maximum at
\begin{equation}
\beta_{max}=0.2656...\;\; .   \label{betac}
\end{equation}
 Together with Eq.(\ref{hypo}) this provides the
transition temperature of the two layer
Ising system. Our preliminary Monte Carlo simulatons,
using Ferrenberg-Swendsen algorithm \cite{fs}, give $\beta_c = .27(1)$.
Obviously higher precision computations are required to decide on
the
discrepancy between these two numbers \cite{my}. At present stage we can
only say that,
even if
Eq.(\ref{hypo}) is not exact, it gives rather good estimate of
$\beta_c$ in this case.
\subsection{Three dimensional Ising model}

  According to Eq.(\ref{rho})
in order to calcuate, say, $\rho_2$ for a $d$ dimensional system
one needs the partition functions of the $d-1$ dimensional model
and that of the two coupled $d-1$ dimensional systems.
 In particular, calculating $\beta_c$ for
the three dimensional Ising
model would be reduced to finding the free energy of the two coupled
layers of
the Ising spins. Indeed in this case the full transfer matrix ${\cal T}$
propagates the whole plane of spins, say, vertically, while the reduced
transfer
 matrices $T_{1,2}$ propagate one (two) rows
horizontally. Analytic solution of this system is not known.
In our previous application
 only the transition temperature was derived. However a simple recursive
  counting of microcanonical states allows to derive exact expressions
for the complete partition functions $Z_{1,2}(\beta,L)$
{\em in the finite volume} for not so large $L$ \cite{bha,stod}.
We have therefore calculated
$\beta_{max}(L)$, the exact location of the maxmum
of the ratio $r_L(\beta)=Z_1(\beta,L)^2/Z_2(\beta,\L)$,
which defines the pseudocritical temperature
at finite volume, c.f. Table 1.
\begin{table}
\caption{Volume dependence of the pseudocritical temperature
for the three dimensional Ising model.}
\center{\begin{tabular}{cc} \hline\hline
 $L$  & $\beta_{max}(L)$   \\  \hline\hline
 $3$  & 0.3317          \\
 $4$  & 0.3067          \\
 $5$  & 0.2938          \\
 $6$  & 0.2859          \\
 $7$  & 0.2698          \\ \hline\hline
\end{tabular}} \newline
\end{table}
The available range of $L$ values is rather
limited, nevertheless one sees
the proper trend in the $L$ dependence.  Our values seem to move, albeit
little slowly, towards
$\beta_c^{Ising3}=.221652(3)$ \cite{has}.
Unfortunately larger sizes are required to
test quantitatively the $L$ dependence against predictions of the finite size
scaling.

\section{Critical equalities for the internal energy}
The maximum principle (\ref{hypo}) can be also stated in another
interesting fashion.
Differentiating
the logarithm of Eq.(\ref{rho}) gives as the condition for the maximum
in the $n=2$ case
\begin{equation}
u_2(\beta_c)=u_1(\beta_c),   \label{us}
\end{equation}
where $u_{1,2}$ denotes the density of the internal energy for the
one (two) layer
system. We use again the three dimensional terminology.
Note slightly unusual normalization of the free energy of a single
layer of spins. It is implied by our definition $Z_1\equiv Tr{\cal T}$.
Periodic boundary conditions in the direction {\em perpendicular}
to the plane
shift the energy density by $-1$ relative to the standard definition.
This trivial modification is nevertheless important for the equality
(\ref{us}).
 Analogous relations follow from applying our ``maximum rule'' to
 the higher moments
of the transfer matrix. They all say that at the bulk  critical
temperature internal energies
of the interacting and noninteracting planes are equal.
This statement follows from duality for the two dimensional
Ising and Potts models with planes replaced by the spin chains.

For infinite $n$ the condition for maximum of $\rho_n(\beta)$
takes yet another very interesting form. In this limit the
original $d$ dimensional system is recovered and we get
\begin{equation}
u_{d}(\beta_c^{(d)})=u_{d-1}(\beta_c^{(d)}) \label{uinf},
\end{equation}
where $u_{d}$ denotes the internal energy density of the $d$
dimensional system, and $\beta_c^{(d)}$ corresponds
to its crtitical temperature. As emphasized before, this equality follows
from selfduality in the case of the two dimensional Ising model.
Alternatively one can check it directly. Indeed
\begin{equation}
u_{2}(\beta_c)=u^{Onsager}_{|2\beta=\log{(1+\sqrt{2})}}=-\sqrt{2}=
u_{1}(\beta_c),
\end{equation}
where
$u^{Onsager}(\beta)=-ctgh(2\beta)\left[1+
{2\kappa^{'}\over \pi} K_1(\kappa)\right]$, $\kappa'=2\tanh{(2\beta)}^2-1$,
$\kappa^2+\kappa'^2=1$, $K_1$ is the elliptic function of the
first kind \cite{huang}, and $u_{1}(\beta)=-\tanh{(\beta)}-1$.
For $d>2$ Eq.(\ref{uinf}) remains unproven similarly
to Eq.(\ref{hypo}).

J. W. thanks B. Bunk for the assistance in filling the last entry in
 Table 1, and A. Sokal for the interesting discussion.

\end{document}